\RequirePackage[T1]{fontenc}
\documentclass[10pt,journal,compsoc]{IEEEtran}

\usepackage{graphicx}%
\usepackage{multirow}%
\usepackage{amsmath,amssymb,amsfonts,amsthm}%
\usepackage{mathrsfs}%
\usepackage{xcolor}%
\usepackage{textcomp}%
\usepackage{manyfoot}%
\usepackage{booktabs}%
\usepackage{algorithm}%
\usepackage{algorithmicx}%
\usepackage{algpseudocode}%
\usepackage{listings}%
\usepackage{tikz}
\usepackage{stfloats}
\usepackage{pgfplots}
\pgfplotsset{compat=1.18} % Pour une meilleure compatibilité des versions
\usetikzlibrary{fillbetween}
\theoremstyle{thmstyleone}%
\newtheorem{theorem}{Theorem}%  meant for continuous numbers
\newtheorem{proposition}[theorem]{Proposition}% 

\theoremstyle{thmstyletwo}%

\theoremstyle{thmstylethree}%

\usepackage{subcaption}

\usepackage{alphalph}

\usepackage[numbers,sort&compress]{natbib}% Citation support using natbib.sty
\bibpunct[, ]{[}{]}{,}{n}{,}{,}% Citation support using natbib.sty
\theoremstyle{plain}% Theorem-like structures provided by amsthm.sty

\usepackage{amsmath,amssymb,amsfonts,amsthm}
\usepackage{hyperref}
\usepackage{orcidlink}
\usepackage{rtsched}
\newcommand{\E}{\mathbf{E}}
\renewcommand{\Pr}{\mathbf{P}}
\newcommand{\dmpig}{\Delta^\text{IG}}

\renewcommand{\epsilon}{\varepsilon}
\renewcommand{\Pi}{\varPi}
\renewcommand{\phi}{\varphi}

\newcommand{\ie}{\textit{i}.\textit{e}.,}

\newcommand{\cf}{\textit{c}.\textit{f}.,}

\newcommand{\simso}{SimSo}
\usepackage{url}
\usepackage{lscape}
\usepackage{cleveref}
\usepackage{booktabs}
\usepackage{xfrac}
\usepackage{alphalph}

\begin{document}

	\title{Response Time Central-Limit and Failure Rate Estimation for Stationary Periodic Rate Monotonic Real-Time Systems}

	%\markboth{~IEEE Transactions on Computers}%
	%{Shell \MakeLowercase{\textit{et al.}}: Bare Demo of IEEEtran.cls for Computer Society Journals}
	
%	\IEEEpubid{0000--0000/00\$00.00~\copyright~2021 IEEE}

\IEEEtitleabstractindextext{%
	\begin{abstract}
			Real-time systems consist of a set of tasks, a scheduling policy, and a system architecture, all constrained by timing requirements. Many everyday embedded systems, within devices such as airplanes, cars, trains, and spatial probes, operate as real-time systems. To ensure safe failure rates (also known as \textit{deadline miss probabilities}), response times must be bounded. Rate Monotonic real-time systems prioritize tasks according to their arrival rate. This paper focuses on the use of the central limit of response times built in \cite{zagalo2022} and an approximation of their distribution with an inverse Gaussian mixture distribution. The distribution parameters and their associated failure rates are estimated through a suitable re-parameterization of the inverse Gaussian distribution and an adapted Expectation-Maximization algorithm. Extensive simulations demonstrate that the method is well-suited for the approximation of failure rates. 
	\end{abstract}
	
	}

\author{Kevin~Zagalo\,\orcidlink{0000-0001-7768-8841} and   Avner~Bar-Hen\,\orcidlink{0000-0002-4449-8117}% <-this % stops an unwanted space
\thanks{This work was entirely funded by INRIA. We gratefully thank Liliana Cucu-Grosjean for her insights.}%
\thanks{Kevin Zagalo is with Inria, INSA Lyon, CITI, UR3720, Villeurbanne, France. E-mail: kevin.zagalo@inria.fr.}%
\thanks{Avner Bar-Hen is with Cnam Paris, 75003 Paris, France. E-mail: avner@cnam.fr.}%
}
	
	\maketitle

	\IEEEraisesectionheading{\section{Introduction} \label{section:introduction}}
	
	\IEEEPARstart{T}{he} increasing demand for new functionalities in embedded systems within automotive, avionics and space industries is driving an increase in the performance required in embedded  processing units \cite{lee2008cyber}. Real-time systems are computing systems in which tasks must meet their time requirements that we refer to as deadlines. If a task does not meet its deadline, it is considered a failure. Embedded systems usually use little energy and computing resources, thus have a specific design with a micro-controller architecture and a set of tasks running on it  \cite{laplante2004real}.  An important part of this design is to associate a processing unit to a given task set. The time taken by the system to respond to an input and provide the output or display the updated information is known as the end-to-end latency. It is composed of two main parts: the communication of the data from the sensors to a task; and the time needed to complete the workload of the task itself called \textit{response time}. In this paper we study response times, and more specifically, we estimate the probability that the response time of a task is larger than a predefined threshold. During run-time, each instance of the tasks competes for processing time on the basis of their priority.  To ensure that every task is executed within its specified timing constraints,  computation resources are allocated to different tasks according to their priority. Depending on the device, missing a deadline can be very costly or even catastrophic. Therefore a classical problem is the study of the performance of the worst-case scenario, \ie~the scenario producing the largest response times. This method is generally efficient but it forces designers to overestimate the quantity of processing units necessary to run a task set. Since the complexity of real-time systems is increasing, the worst-case scenario often becomes unrealistic and over-estimation of Worst-Case Response Time (WCRT) increases drastically. A way to soften resource requirements is to allow a (low) failure rate for each task, such that the probability that a deadline is missed is bounded by this failure rate. %
	%Researchers have suggested using random variables to model execution times.
	%Thus, not only the single value of the worst-case execution time is considered but the whole distribution of execution times. %
	%However, the introduction of statistics and probabilistic models in real-time systems have been focused toward estimating WCRTs as random variables. 

	%The studied scheduling policy in this paper is the preemptive Rate Monotonic (RM) scheduling policy. RM assigns higher priorities to higher rates of arrival: a task will stop (\ie~preempt) any other running task with a lower priority. RM is optimal for real-time systems using static-priorities and implicit deadlines \cite{audsley1991hard}. This means that if RM fails to make a real-time system feasible, then all other static-priority scheduling policies fail.  In \cite{liu1973scheduling} is shown that if the maximum utilization of the system is (asymptotically) lower than $\log(2)$ the system is feasible with failure rates equal to $0$. Hence we focus in this paper on systems with a maximum utilization higher than $\log(2)$.

	\subsection{Related work} 
		%Failure rates (or deadline miss probabilities) are computed to determine if a set of tasks can be scheduled when execution times or arrival times are assumed random. For critical applications, failure rates should be equal to $0$. For less critical applications, failure rates can be tolerated until a threshold given by the application. 
				Historically, the distribution of response times has been computed in a closed form for periodic real-time systems in discrete time with discrete execution times in \cite{diaz2002stochastic}. This approach is based on the convolution operator, and provides an iterative method to compute the exact distribution of response times. However, this convolution operator has a high complexity \cite{markovic2021convolution}. This complexity might not be a problem in the design part of a real-time system, however we aim to approach the timing analysis as part of the scheduling process itself. This complexity is the main reason why the concentration inequalities have been widely used to bound failure rates \cite{chen2018analysis, von2018efficiently, friebe2023continuous, markoviccta, markovic2022analytical,friebe2024efficiently}. 
				
				Failure rates are used in the design process of real-time systems in order to provide quality of service metrics. We propose in this work a first step into the incorporation of the failure rates estimate in the scheduling process itself. It opens the possibility of an adaptive scheduling method taking failure rates into account, by estimating during run-time the distribution and/or the component of the mixture where response times belong. This kind of adaptive algorithm is studied in \cite{purohit2018improving, wei2020optimal, NEURIPS2024_becd02b8} and the impact of such prediction within the scheduling algorithm is studied in \cite{11153169}.  We outline how to build such adaptive algorithms using the proposed approach in Section~\ref{sec:conclusion}.	
							
				The starting point of the failure rate estimation provided in this work is the central limit of response times built in \cite{zagalo2022}, pushing the inverse Gaussian (IG) distribution as a natural candidate to approximate response time distributions. The IG family is a natural choice for the statistical modeling of positive and right-skewed distributions, see \cite{folks1978inverse,tweedie1957statistical}. It is used in many fields, such as industrial degradation modelling, psychology, and many others like hydrology, market research, biology, ecology, etc. \cf~\cite{seshadri2012inverse}. %Modeling of a core system can also be seen from an actuarial point of view: incoming cash premiums and outgoing claims of a classical risk process become incoming task and end of the task. %The ruin corresponds to the moment where the core finished a given task as well as all the task with higher priority and meeting deadline corresponds to have a ruin before a given time \cite{delbaen1987classical,morales2004risk}. 

	\subsection{Contributions}
	The two main contributions are:
	\begin{enumerate}
		\item Theoretical evidence, and extensive simulations, showing that inverse Gaussian mixtures are a suitable distribution family for response times, hence for failure rate estimation,
		\item A practical and simple implementation of backlog estimation using an adapted EM algorithm accounting for the statistics of execution times, in order to estimate failure rates.
	\end{enumerate}
	
	Three methods to approach the failure rates are compared in this paper : empirical  failure rates, a theoretical bound on failure rates and an estimation of the failure rates. Note that only the theoretical bounds can determine safely the feasibility of a real-time system, \ie~failure rates equal to $0$. We emphasize  the fact that the estimated failure rates should not in any case be considered as a measure of reliability of the system. In Section~\ref{sec:RT}, stationary rate monotonic real-time systems are introduced. A theoretical bound on failure rates (the Hoeffding bound) is provided as a baseline. We then propose a suitable parameterization of the IG distribution in Section~\ref{sec:RTIG} adapted to response times, in the form of \eqref{eq:mixture}. This reparametrization is based on the central limit-like result given in \cite{zagalo2022}. Using an adapted EM algorithm, we estimate the parameters of a mixture of IG probability density functions (p.d.f.). Failure rates are estimated through this parameterization.	Finally in Section~\ref{sec:simulations} we apply the foregoing method to simulated and real data. All proofs are provided in appendices. %Most of the plots are given in Appendix~\ref{plots}.%, and discuss the perspectives of such method in real-time systems.

	\section{Stationary periodic rate monotonic real-time systems}\label{sec:RT}
	
	In this paper we consider \textit{periodic} tasks, meaning that an instance of each task is periodically released at a given rate, and a single core system, \ie~only one task is processed at a time. Hence a real-time system is referred to only with its task set. The deadlines are \textit{implicit}, \ie~an instance of a task should finish its execution before the next instance of the same task in order to respect its timing requirements. Note that all notations are summarized in Table~\ref{tab:notations}.
	
	\begin{table*}[t]
	\caption{Notations used throughout the paper.}\label{tab:notations}
	\medskip
		\begin{tabular}{l|p{0.38\textwidth}|l|p{0.42\textwidth}}
%	$\Pr$ & probability measure& $\E$ & expectation operator\\
	$\Gamma$ & task set&$\gamma_i$ & task of index $i$\\
	$\gamma_{i, j}$ & $j$-th job of the task of index $i$&$\Delta_i$ & failure rate of the task of index $i$\\
	$\Delta_i^{(n)}$ & empirical failure rate of the task of index $i$&$\dmpig_i$ & approximated failure rate of the task of index $i$\\
	$C_i$ & execution time of the task of index $i$ &$C_{i,j}$ & execution time of the $j$-th job of the task of index $i$\\
	$f_i$ & p.d.f. of the execution time of the task of index $i$&$u_i$ & mean utilization of level $i$ of $\Gamma$\\
	$u_i^{max}$ & maximum utilization of level $i$ of $\Gamma$&$v_i$ & standard deviation of level $i$ of $\Gamma$\\
	$v_i^{max}$ & maximum deviation of level $i$ of $\Gamma$ &$R_{i, j}$ & response time of the $j$-th job of the task of index $i$ \\
	$R_{i}$ & response time of the mixture of all jobs of the task of index $i$ &$h_i$ & p.d.f. of the response time of the mixture of all jobs of the task of index $i$   \\
	$A_{i, j}$ & arrival time of the $j$-th job of the task of index $i$&$N_i$ & counting process of the task of index $i$\\
	$O_i$ & offset -arrival time of the first job- of the task of index $i$&$p_i$ & time separating two consecutive jobs of the task of index $i$\\
	$\lambda_i$& rate of arrival of the jobs of the task of index $i$&$\beta_{i, j}$ & backlog of jobs of index strictly lower than $i$ at time $A_{i, j}$ \\
	$\hat \beta_{i, j}$ & approximation of $\beta_{i, j}$ &$m_{i, j}$ & distribution of the backlog of jobs of index strictly lower than $i$ at time $A_{i, j}$ \\
	$\pi_i$ &  distribution of the backlog of index strictly lower than $i$&$\pi_{i, k}$ & weight of the $k$-th component of the mixture of the response time of the task of index $i$\\
	$\hat \pi_{i, k}$& approximation of $\pi_{i, k}$&$K_i$ & number of components of the IG mixture for task $i$\\
	$\psi(r ; \xi, \delta)$ & p.d.f. of the IG distribution of mean $\xi$ and shape $\delta$&$\mu_i(\beta)$ & mode of  the IG distribution of mean $\frac{\beta}{1-u_i}$ and shape $\frac{\beta^2}{v^2_i}$\\
	$\tilde\psi(r; \mu, \nu)$ & p.d.f. of the reparametrized IG distribution of mode $\mu$ and variation coefficient $\nu$&$\nu_i$ & variation coefficient of the IG distribution of mean $\frac{\beta}{1-u_i}$ and shape $\frac{\beta^2}{v^2_i}$\\
	$\Psi(r; \xi, \delta))$ & c.d.f. of the IG distribution of mean $\xi$ and shape $\delta$ &	$\Phi$ & c.d.f. of the $\chi^2(1)$ distribution\\

\end{tabular}
\end{table*}

		\subsection{Model} \label{sec:model}
	Let us consider a periodic real-time system $\Gamma = (\gamma_i)_i$. Without loss of generality, the task indices are ordered by decreasing priority: If $k<i$, $\gamma_k$ preempts $\gamma_i$ if both $\gamma_i$ and $\gamma_k$ are activated. Task prioritization ensures that the most critical and time-sensitive tasks are completed first to meet deadlines. A task $\gamma_i$  is characterized by: 
	\begin{enumerate}
		\item its \textit{execution time} $C_i$,  with a p.d.f.  $f_i$ of non-negative and finite support $[c_i^{min}, c_i^{max}]$.
		\item its \textit{period} $p_i > 0$. We note $\lambda_i = 1/p_i$ the rate of arrival of $\gamma_i$. $p_i$ is also its \textit{deadline}, \ie~the maximum time given to the task $\gamma_i$ to be processed.
		%\item its \textit{permitted failure rate} $\alpha_i \in [0,1)$.
	\end{enumerate}
	
	We suppose that the execution times of the jobs and tasks are statistically independent and independent of the arrival offsets. All results strongly depend on this assumption.
	
	Let us first introduce some concepts that will be used throughout this paper.

	\subsubsection{Utilization}Let us define \textit{$k$-level mean utilization} of $\Gamma$ as $u_k = \sum_{i=1}^k \lambda_i \E[C_i]$. The \textit{$k$-level mean utilization} takes into account all tasks with higher or equal priority than the task $\gamma_k$. We also define the \textit{$k$-level maximum utilization} of $\Gamma$  as $u_k^{max}= \sum_{i=1}^k  \lambda_i c_i^{max}.$
	
	\subsubsection{Deviation}Let us define the \textit{$k$-level deviation} of $\Gamma$ as $v_k =\left(\sum_{i=1}^k \lambda_i \E[(C_i-\E[C_i])^2]\right)^{\sfrac{1}{2}}.$ The \textit{$k$-level  deviation} takes into account all tasks with higher or equal priority than the task $\gamma_i$. We also define the \textit{$k$-level maximum deviation} of $\Gamma$  as $v_k^{max}= \left(\sum_{i=1}^k \lambda_i (c_i^{max}-c_i^{min})^2\right)^{\sfrac{1}{2}}.$

	\subsubsection{Jobs}
	A \textit{job} $\gamma_{i,j}$ is the $j$-th instance of the task $\gamma_i$. $C_{i,j}$ is its execution time of p.d.f. $f_i$ and $A_{i,j}$ is its arrival time such that \begin{equation}\label{eq:arrival}
		\forall j \geq 1, \quad A_{i, j+1} - A_{i,j} = p_i,
	\end{equation} which implies that, the time between two consecutive jobs of $\gamma_i$ is $p_i$, with probability $1$.  Note that the periodicity of tasks is purely chosen in order to keep the implementation and proof simple. What is strictly necessary for the results to hold is the \textit{stationarity} of the system.

	\subsubsection{Stationarity}
	We assume the \textit{stationarity} of $\Gamma$, that we define as the stationarity of all demand processes $W_i(t) = \sum_{j=1}^{N_i(t)} C_{i,j}, $ where $N_i(t) = \sum_{j=1}^\infty \mathbf{1}_{[0,t]}(A_{i, j})$ is the process counting the jobs of the task $\gamma_i$ arrived before instant $t\geq0$. This means that we assume that the distribution of the number of jobs arrived in an interval $[s,t]$ is the same as  the one over any time interval $[u , u + t-s]$ of size $t-s$, \ie~$\Pr(N_i(t) - N_i(s) = n) = \Pr(N_i(t-s) = n)$. Since the arrival process $A_i = (A_{i, j})_j$ verifies \eqref{eq:arrival},  it is easy to verify  that it suffices that the first arrival (also known as offset) $O_i = A_{i, 1}$ is continuously and uniformly distributed between $0$ and $p_i$.  Note that  $N_i(t) = \lceil \lambda_i (t - O_i)^+ \rceil$\footnote{where $x^+ = \max(0, x)$.} and $\E[N_i(t)] = \lambda_i t$, and that $\sum_i N_i$ is not stationary in general. From the stationarity of $N_i(t)$ and the fact that the $(C_{i,j})_j$ are i.i.d. implies that $W_i(t+s) - W_i(s)$ has the distribution of $W_i(t)$, i.e. $W_i$ is stationary. The sum $\sum_i W_i$ is not stationary in general either.  Stationarity is central to the proposed analysis (see Proposition~\ref{prop:centrallimit}), making the proposed estimation independent of the arrival time of jobs. In a non-stationary setup, the arrival time $A_{i,j}$ should be taken into account in the proposed estimation, and the central-limit of Proposition~\ref{prop:centrallimit} does not hold in general. 
	
	\subsubsection{Backlog}
		In order to make the analysis simpler, we will make the assumption that whenever a job misses its deadline, it is considered useless and is discarded from the system. Such discarding policy is used in systems such that, after the deadline of a given job, its content is useless to the system. This will ensure that the worst-case scenario is the simultaneous activation of all tasks, in the sense that it provides the largest response time.  The backlog of the task $\gamma_i$ is defined as the amount of time required at a given time to terminate all active jobs of the task $\gamma_i$. The backlog process of a task $\gamma_i$ can be defined as $\beta_{i}(t) = C_{i,N_i(t)} - \int_{A_{i,N_i(t)}}^t \!\!\!\! \mathbf{1}(\beta_i(s) > 0)\prod_{k =1}^{i-1} \mathbf{1}(\beta_{k}(s) = 0) ds$. The backlog process was first introduced in real-time system timing analyses in \cite{diaz2002stochastic} in the discrete case with no discarding, and \cite{zagalo2022} in the continuous case, where it is shown that this backlog process is ergodic, if and only if $u_i < 1$, \ie~it converges with probability $1$. The discrete time analysis provided in \cite{diaz2002stochastic} induced a literature based on the convolution operator. In the discrete time case, the distribution of $\beta_{i}(A_{i, j+1})$ has been extensively studied using the convolution of the p.d.f. of each individual execution time at each arrival time, as well as the distribution of the limit $\lim_{t\to\infty}\beta_{i}(t)$. Nevertheless, the workload that needs to be terminated before a job $\gamma_{i,j}$ is $\sum_{k=1}^{i-1} \beta_k(A_{i,j})$. Unfortunately, the cumulated backlog process $\sum_{k=1}^{i-1} \beta_k$ is not stationary in general.

	\subsubsection{Response times}
		 The \textit{response time} of $\gamma_{i,j}$ is denoted $R_{i,j}$, and is the time elapsed between $A_{i,j}$ and the end of its execution. Thus a response time of a job is the sum of the execution time of this job and the execution times of higher priority jobs that preempt it, see \figurename~\ref{exSchedule}. Hence, it depends on the scheduling policy. For the discarding policy, $R_{i,j}$ denotes the virtual completion time obtained by retaining the job until completion, keeping the higher-priority tasks under their original discard policy, and preventing later jobs of the same task from delaying it. It agrees with the actual response time whenever the job completes by its deadline. Thus 
		 \begin{equation}\label{def:responsetime}
		 	R_{i,j} = \inf\left\{t>0:
             \int_{A_{i,j}}^{A_{i,j}+t}\prod_{k=1}^{i-1}
             \mathbf{1}(\beta_k(s)=0)\,ds\geq C_{i,j}\right\}.
		 \end{equation}
		 
		 The relative deadline of a task, \ie~its maximum allowed response time, is equal to its period. This means that the response time $R_{i,j}$ of the job $\gamma_{i,j}$ should be lower than $p_i$. Therefore, the absolute deadline of $\gamma_{i,j}$ is $A_{i,j+1}$. Note that the response times are dependent on the scheduling policy used with the system. In this paper we use the RM scheduling policy that we define in the following section.  A job is discarded if it misses its deadline, and with implicit deadlines there can be at most one active job per task in the actual schedule. The actual deadline-miss event is $R_{i,j}>p_i$, since the actual and virtual schedules agree up to the deadline. A discarded job provides only this inequality; fitting its virtual response time requires either reconstruction or an estimation method that accounts for censoring. 
		 
		 The parametric estimation provided in this work relies on the convergence in distribution built in \cite{zagalo2022}. 
		 
		 \begin{proposition}[Central limit]\label{prop:centrallimit}
				Given a backlog $\sum_{k=1}^i \beta_k(A_{i,j}) = \beta$, the response time $R_{i,j}$ converges in distribution to an inverse Gaussian distribution when the mean utilization $u_i$ goes to $1$ from below, i.e. 
				\begin{equation}\label{eq:centrallimit}
				\Pr(R_{i,j} \leq t \mid\!\sum_{k=1}^i \beta_k (A_{i,j}) = \beta) \!\! \underset{u_i \to 1^-}{\simeq} \!\!\! \Psi\!\left(t; \dfrac{\beta}{1-u_i},  \dfrac{\beta^2}{v_i^2} \right),
				\end{equation}
				where $\Psi(r; \xi, \delta)$ is  the c.d.f. of the inverse Gaussian distribution of mean $\xi$ and shape $\delta$.
		 \end{proposition} 
	
	\begin{proof}
	From Lemma~1 and 2 in \cite{zagalo2022}, we get that, with $C_{i,j} = x$ and $\sum_{k=1}^{i-1}\beta_k(A_{i,j}) = y$, $R_{i,j}$ converges to an inverse Gaussian variable of mean $\dfrac{x+y}{1 - u_i}$ and shape $\dfrac{(x+y)^2}{v_i^2}$ when $u_i$ gets close to $1$. We conclude with the fact that $\sum_{k=1}^{i}\beta_k(A_{i,j}) = C_{i,j} + \sum_{k=1}^{i-1}\beta_k(A_{i,j}). $
	\end{proof}
	
	\subsubsection{Failure rate} 
	The \textit{failure rate} of the task $\gamma_i \in \Gamma$ is defined by \begin{equation}\Delta_i = \lim_{n\to\infty} \dfrac{1}{n} \sum_{j=1}^n \Pr\left(   R_{i,j} > p_{i} \right).\label{eq:failurerate}\end{equation} Thus finding an analytical expression of the p.d.f. of response times permits to determine the failure rates. Moreover, in \cite{zagalo2022}, it is proven that in the case of stationary real-time systems, the sequence $(R_{i, j}, j \in \mathbb{N})$ is stationary. Consequently, $\Delta_i=\Pr(R_{i,j}>p_i)$ is independent of $j$. Stationarity does not imply a finite mixture representation. The number of components $K_i$ is introduced as a modeling choice in \eqref{eq:mixture} and selected in Section~\ref{sec:bic}.

	\subsubsection{Rate Monotonic}
		\begin{figure}[t]
		\centering
		\begin{RTGrid}[width=0.92\linewidth]{3}{19}  
			\TaskNArrDead{1}{0}{4}{4}{4}    % draws the arrivals and deadlines
			\TaskNExecDelta{1}{0}{1}{4}{5}  % draws executions (highest priority) 
			\TaskNEnd{1}{1}{4}{5}
			\TaskNEnd{2}{4}{6}{3}
			\TaskNArrDead{2}{0.2}{6}{6}{3}   % draws the arrivals and deadlines
			\TaskRespTime{2}{0.2}{3.8}         % draws the hatched rectangle in [0,4]
			\TaskExecution{2}{1}{4}        % draws execution (over the previous rectangle)
			\TaskRespTime{2}{6.2}{4}         % draws the hatched rectangle in [6,10]
			\TaskExecution{2}{6.2}{8}        % draws execution
			\TaskExecution{2}{9}{10.2}       % draws execution
			\TaskRespTime{2}{12.2}{3.8}        % draws the hatched rectangle in [12,16]
			\TaskExecution{2}{13}{16}      % draws execution 
			
			\TaskNEnd{3}{18}{18}{1}
			
			\TaskNArrDead{3}{1.7}{18}{22.7}{1}   % draws the arrivals and deadlines
			\TaskRespTime{3}{1.7}{16}         % draws the hatched rectangle in [0,4]
			\TaskExecution{3}{5}{6.2}        % draws execution (over the previous rectangle)
			\TaskExecution{3}{10.2}{12}        % draws execution
			\TaskExecution{3}{17}{18}        % draws execution
		\end{RTGrid}
		\caption{Example of a static-priority schedule with $O_1 =0, C_1 = 1, p_1=4, O_2=0.2, C_2=3, p_2=6, O_3=1.7, C_3=4, p_3=18$. The higher priority tasks stop the execution of tasks if needed. Thus, the response time is time between the release and the end of a job, taking the jobs of higher priority tasks into account. Corner dots represent the end of a job, up-arrows their arrival time, down arrows their absolute deadlines.} \label{exSchedule}
	\end{figure}

	We consider a real-time system $\Gamma$ ordered by decreasing priority order and scheduled with the RM policy: $\Gamma$ is such that priorities are ordered decreasingly with the rates, \ie~$\lambda_{i} \geq \lambda_{i+m}$ for any $m \geq 0$. RM is optimal for real-time systems using static-priorities and implicit deadlines \cite{audsley1991hard}. By optimal, we mean that if RM fails to schedule $\Gamma$, then all static-priority scheduling policies also fail to make $\Gamma$ feasible.  If $u_i>1$, the corresponding non-discarding workload cannot be stable under the usual workload assumptions; this does not imply $\Delta_i=1$ under discarding. The seminal work of Liu and Layland \cite{liu1973scheduling} provides a sufficient condition for the feasibility of any system with finite supports of execution times using the \textit{$k$-level maximal utilization}.  Whenever \begin{equation} \label{eq:LL} u^{max}_k <k ( 2^{\sfrac{1}{k}} - 1)  \end{equation}  the task set $\Gamma$ is such that for all $i \leq k$, and all $j \in \mathbb{N}$, the probability of a failure $\Pr(R_{i,j} > p_i)$ is equal to zero \cite[Theorem 5]{liu1973scheduling}, and thus $\Delta_1 = \dots = \Delta_k = 0$. Moreover, while $u_k^{max} < 1$ there exists a scheduling policy, with dynamic priorities that can satisfy the feasibility of the task $\gamma_i$ \cite{liu1973scheduling}. These utilization thresholds distinguish sufficient schedulability conditions from regimes in which deadline misses may occur; they are not universal phase transitions. The Liu--Layland threshold decreases to $\log(2)$ as the number of tasks increases, and exceeding it does not by itself imply a deadline miss. As proven in \cite{diaz2002stochastic}, the necessary condition for the feasibility of a task $\gamma_k$ is that the mean utilization $u_k$ is less than one. Hence, there is a room for an improvement between the necessary condition provided in \cite{diaz2002stochastic}, \cf~$u_k < 1$ and the sufficient condition $u^{max}_k < k ( 2^{\sfrac{1}{k}} - 1)$. In particular in the case where $u_k < 1$ and $u^{max}_k > 1$ as we see in \figurename~\ref{fig:dmp}.

	\subsection{Hoeffding bound for stationary periodic rate monotonic real-time systems} \label{sec:Hoeffding}
	
	The critical instant approach in the deterministic case has been used in the probabilistic setup for several years, and has been refuted since then in \cite{chen2022critical}, where authors explore a counter example, and provide solutions to bypass the issue: one of those considers adding a job per task to the timing analysis. This method is called the \textit{carry-in} method, which is used in the proof of the following proposition (see Appendix~\ref{proof:hoeffding}). Thus, we derive the Hoeffding bound associated to the carry-in method as a state-of-the-art baseline.  
	
	\begin{proposition}[Hoeffding bound] \label{HB}
        Let $\Gamma$ be a stationary periodic rate-monotonic real-time system as defined in Section~\ref{sec:model}, $\gamma_i\in\Gamma$, $i>1$, and suppose $u_i<1$, $u_i^{max}>i(2^{1/i}-1)$ and $v_i^{max}>0$. If $p_i\geq\dfrac{3\sum_{k=1}^i\E[C_k]}{1-u_{i-1}}$, then
        $$\Delta_i\leq\exp\left[-\frac{2\lambda_i}{3} \left(\frac{\sum_{k=1}^i\E[C_k]}{v_i^{max}}\right)^2\right].$$
        %If $v_i^{max}=0$, the same hypothesis implies $\Delta_i=0$.
    \end{proposition}
	
	\begin{proof}

			Let us consider a task $\gamma_i$ where $i > 1$. Firstly let us remark that $R_{i,j}$ is upper-bounded with probability $1$ by $R_{i, j} \leq \inf\{ t > 0 : \sum_{k=1}^i \beta_{k}(A_{i,j}) + W_{i-1}(t + A_{i,j}) -  W_{i-1}(A_{i,j}) \leq t \}$. Thanks to the discarding policy, there cannot be more than one job per task activated at the same time. Thus, the tail function of $R_{i, j}$ is upper-bounded by the tail function of the RHS. From the stationarity of all demand processes, and the Hoeffding inequality \cite{Hoeffding01031963}, we get the upper-bound 					 	
			\begin{multline}\label{eq:hoeffding} 
				-\ln\Pr(R_{i,j} > t \mid O_1, \dots, O_i)  \\
							\leq 2  \!\frac{(t- \!\sum_{k=1}^{i} \!\E[ C_k] + \sum_{k=1}^{i-1} \!\sum_{l=N_k(A_{i,j})}^{N_k(A_{i,j} + t)}  \!\E[C_{k,l}]])^2}{\sum_{k=1}^i  (c_k^{max} - c_k^{min})^2 (N_k(A_{i,j} + t) - N_k(A_{i,j}))}.
			\end{multline}		
			For any $t \geq 0$, $N_{i-1}(t)$ satisfies $N_{i-1}(t) \leq \lambda_{i-1} t + 1$. Since only one job of $\gamma_i$ is released before $p_i$, we have $W_{i-1}(t) = \sum_{k=1}^{i-1} \sum_{j=1}^{N_k(t)} C_{k,j}$ which yields $\E[W_{i-1}(t) \mid O_1, \dots, O_{i-1}] =  \sum_{k=1}^{i-1}N_k(t)\E[C_k] \leq   u_{i-1}t + \sum_{k=1}^i \E[C_k]$.  Hence, for $u_{i-1} < 1$, we get $u_{i-1} t + 2\sum_{k=1}^i \E[C_k] \geq \E[W_{i-1}(t) +  \sum_{k=1}^i C_k \mid O_1, \dots, O_{i-1}],$ which implies that, $t > \E[W_{i-1}(t) +  \sum_{k=1}^i C_k \mid O_1, \dots, O_{i-1}]$ if  $t \geq \frac{2\sum_{k=1}^i \E[C_k]}{1-u_{i-1}}$.  We are interested in the case $t = p_i$, yielding			
			\begin{multline}
				\frac{(p_i-\E[W_{i-1}(p_i)+ \sum_{k=1}^{i} C_k  \mid O_1, \dots, O_{i-1} ])^2}{\sum_{k=1}^i (c_k^{max}-c_k^{min})^2 (N_k(p_i)+1)} \\ \geq \frac{p_i^2(1- u_{i-1} - 2\lambda_i\sum_{k=1}^{i}\E[ C_k] )^2}{ \sum_{k=1}^i (2+ \lambda_kp_i )(c_k^{max}-c_k^{min})^2}.  \label{eq:inequality}
			\end{multline}

			Since we are using the RM policy,  we have $\lambda_k \geq \lambda_i$ for $i \geq k$, hence we get $\sum_{k=1}^i (1+\lceil\lambda_k p_i\rceil)(c_k^{max}-c_k^{min})^2 \leq 3 p_i (v_i^{max})^2$. Furthermore, $1-u_{i-1} - 2\lambda_i\sum_{k=1}^{i}\E[ C_k] > \lambda_i\sum_{k=1}^{i}\E[ C_k]  $ which gives us the result with  the fact that $x \to \exp(-x)$ is decreasing. We conclude by noticing that the RHS in \eqref{eq:hoeffding} does not depend on $j$, hence it holds for all jobs of the task $\gamma_i$.
		
	\end{proof} 
	Note that the Hoeffding bound approach has  been used in \cite{von2018efficiently} in a different context. We use the Hoeffding bound to compare it to the results of the estimation built in the next sections. Indeed, the failure rate estimation provided in this work are not guaranteed, they are estimations that could be for example incorporated in a predictive scheduling algorithm \cite{purohit2018improving}. We do not investigate such algorithm, but provide a well defined and parametrized method to make one.

	%\color{red} Such bound is practical to establish the schedulablity of the system, however 
	%there exists a time $t^*_i$ such that the response times of jobs of priority level $i$ released after $t^*_i$ are stationary, \ie~the sequence $\{ R_{i,j} : A_{i,j} > t^*_i\}$ is stationary. Before $t^*$ the system is said transient. %For jobs released before $t^*$, the worst-case response time $\sup_j R_{i,j}$ is considered, and its distribution is also found in \cite{zagalo2022}.  We treat in this paper the transient and stationary response time  and estimate its failure rate. 
	
	\section{Inverse Gaussian distribution and response times}\label{sec:RTIG}
	
	The arrival of jobs can be represented as a queueing process with a deterministic activation rate (D), a general execution time distribution (G), and one server (1) since we consider a single process scheme with fixed priorities (FP). This model is widely used in the analysis of various types of service systems such as call centers, banks, or hospitals and has been widely studied \cite{baccelli2013elements}. An approximation of response times is determined by using the \textit{heavy-traffic assumption} \cite{lehoczky1996real, zagalo2022} on the D/G/1/FP queues associated to the point processes $(A_i)_i$ ordered by priority, and the execution time distributions which have a finite support, which permits to approximate the size of a queue with a Brownian motion. 
	
	In \cite{zagalo2022}, authors propose to approximate response time distributions in the fixed-priority setup, with inverse Gaussian distributions, by conditioning the backlog process values. They provide a result analogous to a central limit theorem on response times, using heavy-traffic theory \cite{lehoczky1996real} : if a job of priority $i$ starts with a backlog $\beta$, the p.d.f. of its response time tends to\begin{equation} \psi(r;\xi, \delta) =  \sqrt{\dfrac{\delta}{2\pi r^{3}}} \exp\left(-\dfrac{\delta(r-\xi)^2}{2r\xi^2} \right), \quad r \geq 0\label{eq:inverseGaussienne} \end{equation} when $u_{i}$ tends to $1$ from below, and where $\psi(r ; \xi ,\delta)$ is the p.d.f. of the inverse Gaussian distribution of mean $\xi$ and shape $\delta$. In this paper, we provide an approximation of the failure rates associated to this central limit in the case of a rate monotonic scheduling policy. % The IG p.d.f.  $\psi$ is  defined by  where $\xi$ corresponds to the mean and $\delta > 0$ is called the shape.% More precisely they proved that for $k,$ such that $a_{k,l}\leq A_{i,j}$ (\ie tasks before $R_{i,j}$), $R_{i,j}|\gamma_{k,l}\sim \psi(x ;\xi=\beta / (1-u_{i}), \delta=\left(\beta / v_{i}\right)^{2})$. 
	The key parameter $\beta$ is directly related to the arrival of the previous tasks and can be estimated from the empirical data.  In \cite{zagalo2022}, authors prove that when the mean utilization $u_i$ is smaller than $1$, the distribution of response times can be approximated by a mixture of IG distributions \eqref{eq:firstapproximation}, \ie~ they provide a central limit theorem for response times, \begin{equation}\label{eq:firstapproximation} \dfrac{\Pr(R_{i,j} \!\in\! [r, r+dr])}{dr}\! \!\!\!\!\underset{u_i \to 1^-}{\simeq} \!\!\int_0^\infty \!\!\!\!\psi \left(\!r ; \!\dfrac{\beta}{1-u_i}, \dfrac{\beta^2}{v^2_i} \!\right)\!m_{i,j}(d\beta), \end{equation} where $m_{i,j}$ is a probability measure on backlogs that we consider unknown. The purpose of finding the failure rate is to quantify the quality of the scheduling. Empirical measurements  of response times however are not sufficient, as they are only a fraction of the possible values. Inference methods are used in the purpose of quantifying failure rates from measurements taking into account the shape that response times distributions should have. Nevertheless, the estimation method provides no strict bound, as the Hoeffding bound (Proposition~\ref{HB}) does for example, but rather gives a scale of what to expect in terms of failure rates when using inference methods coupled with measurement-based methods.
	%The quality of the estimation is then measured with % to how close the estimation is to measurements but also they must be bigger. 
	
	% \begin{figure}[t]
		%        \centering
		%        \includegraphics[width=\textwidth]%{figures/worstcase_stationary_task_5.eps}
		%        \caption{Example of the MLE of the transient response time, and the stationary and worst-case response time distribution.}
		%        \label{fig:transient}
		%    \end{figure}

	\subsection{Response time IG mixture model}
	
	In this section, we derive the foregoing central limit using the approximation in \eqref{eq:firstapproximation}. Consider a stationary periodic rate monotonic real-time system $\Gamma$, and for $\gamma_i \in \Gamma$ suppose $u_i < 1, u^{max}_i > i(2^{1/i} - 1)$. Let $R_i$ be the variable of p.d.f. \begin{equation}h_i(r) = \lim_{n \to \infty} \dfrac{1}{n} \sum_{j=1}^n \dfrac{\Pr(R_{i,j} \in [r, r+dr])}{dr}, \label{eq:pdf} \end{equation} such that $\Delta_i = \int_{p_i}^\infty h(r)dr$. Using both \eqref{eq:firstapproximation} and \eqref{eq:pdf} gives the IG approximation
	
	\begin{equation}\label{eq:approximation}
		h_i(r) \underset{u_i \to 1^-}{\simeq} \int_0^\infty \psi \left(r ; \dfrac{\beta}{1-u_{i}}, \dfrac{\beta^2}{v^2_{i}} \right)\pi_{i}(d\beta),
	\end{equation} where the stationary distribution of backlogs $\pi_i =  \lim_{n \to \infty} \dfrac{1}{n} \sum_{j=1}^n m_{i, j}$ exists and is unique according to \cite{zagalo2022} in the continuous time case, and \cite{diaz2002stochastic} in the discrete time case.  We approximate the stationary backlog measure $\pi_i$ by a discrete measure with $K_i$ components. This modeling approximation  gives the finite sum  \begin{equation} \label{eq:mixture} h(r ; \boldsymbol{\pi}_i,\boldsymbol{\beta}_i)= \sum_{k=1}^{K_i} 
		\pi_{i, k} \psi\left(r ; \dfrac{\beta_{i, k}}{1-u_{i}}, \dfrac{\beta_{i,k}^2}{v^2_{i}} \right) ,
	\end{equation}
	by finding the appropriate parameters $(\boldsymbol{\pi}_i,\boldsymbol{\beta}_i, K_i)$, which we find by maximizing the likelihood of an IG mixture model. Hence, we get an approximation of the failure rate $\Delta_i$, namely \begin{equation}\label{eq:dmpig}\dmpig_i =  \int_{p_i}^\infty h(r ; \boldsymbol{\pi}_i,\boldsymbol{\beta}_i)dr.
	\end{equation}	By expanding \eqref{eq:dmpig}  we get the following approximated failure rate, 
	\begin{equation}\label{eq:dmp} 
	\dmpig_i = \sum_{k=1}^{K_i} \pi_{i,k} \left[1 -\Psi\left(p_i ; \dfrac{\beta_{i, k}}{1-u_{i}}, \dfrac{\beta_{i,k}^2}{v^2_{i}} \right) \right],	\end{equation} 
	
	where we estimate $(\boldsymbol{\pi}_i,\boldsymbol{\beta}_i, K_i)$ in the following sections.
	%  For example, a task $\gamma_i$ should not miss its deadline with a failure rate $\Delta_i$, and the approximation in \eqref{eq:approximation} is estimated with the mixture \eqref{eq:mixture}.
	
	\subsection{Re-parameterized IG distribution for response times} 
	
	% As the representation in \eqref{eq:responsetimesallsched} is helpful to real-time designers, the parameters both depend on the blocking time distribution. 
	%In order to make a better estimation and more powerful tests, we re-parameterize the IG distribution. 
	The purpose of this section is to provide the efficient distribution family for an approximation of response times and an adapted EM algorithm to estimate the parameters of this approximation. This adapted re-parameterization of the IG distribution reduces the number of parameters of the mixture model. This increases the speed of convergence of the estimation algorithm as well as the stability of the estimates. Furthermore, as underlined in \cite{punzo2019new}, the log-likelihood of the IG distribution has flat regions thus the EM algorithm has very tiny variations. Reducing the number of parameters addresses a part of this problem. A second reduction of this problem is the use of the Aitken acceleration procedure \cite{aitken1926iii}.
	
	In \cite{punzo2019new}, the author introduces a modified version of the IG distribution of parameters $(\xi, \delta)$, using its \textit{variability coefficient}  $\nu = \xi^2 / \delta$ and its mode $\mu = (\xi^2 + 2.25 \nu^2)^{\sfrac{1}{2}} - 1.5\nu$ instead of the shape and the mean. This re-parameterized IG distribution (rIG), $\tilde{\psi}$, of parameters $(\mu, \nu)$ is defined by the probability density function $\tilde{\psi}(r ; \mu, \nu) = \sqrt{\frac{\mu(3\nu+\mu)}{2\pi \nu r^3}} \exp\left\{ - \frac{\left(r - \sqrt{\mu(3\nu+\mu)}\right)^2}{2\nu r}\right\}$. %so that $\psi(r; \mu, \nu) = \psi(r; \xi, \nu)$ when $\mu = \xi \left( \sqrt{1 + \dfrac{9\xi^2}{4\nu^2}} - \dfrac{3\xi}{2\nu} \right)$ and $\nu = \xi^2/\nu$.
	
	% and allows the clustering of the EM-algorithm to be more intuitive and in the case of real-time systems
	
	% Its maximum likelihood estimator is found by the \textit{Expectation-Maximization} (EM) algorithm described in \cite[Section 3.3.2]{punzo2019new}. 
	
	With the rIG distribution applied to the form that take the parameters of the distributions of response times, one can see that only the mode is sensitive to the mixture provided in \eqref{eq:mixture}.
	The variability coefficient of an IG distribution of mean $\beta /(1-u_{i})$ and shape $\left(\beta /v_i \right)^{2}$ is $\nu_i = \dfrac{v_i^2}{(1-u_i)^2}$ and its mode is $\mu_i(\beta) =  \sqrt{\left(\dfrac{\beta}{1-u_i}\right)^2 + \dfrac{9\nu_i^2}{4}} - \dfrac{3\nu_i}{2}$ so that $\psi_i(x ; \beta) = \tilde{\psi}(x ; \mu_i(\beta), \nu_i)=\psi \left(r ; \dfrac{\beta}{1-u_{i}}, \dfrac{\beta^2}{v^2_{i}} \right)$ is the probability density function of an rIG distribution of mode $\mu_{i}(\beta)$ 
	and variability $\nu_{i}$. Therefore the variability of the component of the mixture of IG distributions does not depend on the backlog $\beta$.

	\subsection{Expectation-Maximization : estimating the backlog} \label{sec:mle}

	\begin{figure}[t]
		\centering
		
		\begin{tikzpicture}[xscale=1, yscale=0.9]
			% Toutes les options de l'axe SONT ICI, sur une seule ligne.
			\begin{semilogyaxis}[xlabel={Mean utilization}, ylabel={Failure Rate}, 	width=\linewidth,		height=5.5cm,
xmin=0, xmax=95, ymin=1e-6, ymax=1e0, xtick={0, 20, 40, 60, 80}, xticklabel=\pgfmathprintnumber\tick\%, ytickten={-6,-5,-4,-3,-2,-1,0}, legend pos=north west, legend style={fill=white, draw=black,legend cell align=left}, extra x ticks={40, 55}, extra x tick style={grid=major, dotted, draw=black}, extra x tick labels={}]
				
				% --- 1. Hoeffding bound (Noir) ---
				\addplot[black, thick, smooth] 
				coordinates {
					(20, 1e-6) (30, 1e-4) (40, 1.5e-2) (50, 6e-2) (60, 1e-1) (70, 3e-1) (80, 6e-1) (90, 8.5e-1)
				};
				\addlegendentry{Hoeffding}
				
				% ----------------------------------------------------------------------
				% --- 2. Delta^IG (Orange) : Remplissage Manuel (Coordonnées corrigées) ---
				
				% Remplissage de la bande Orange : Tracé Supérieur + Tracé Inférieur Inversé
				\addplot[orange!30, fill, opacity=0.5, draw=none, forget plot, smooth] 
				coordinates {
					% Courbe Supérieure (IGupper)
					(45, 1e-5) (50, 1e-4) (55, 1e-3) (57, 5e-4) (60, 5e-3) (65, 3e-2) (70, 4e-2) (80, 2.5e-1) (90, 6e-1)
					% Courbe Inférieure (IGlower) en SENS INVERSE (Corrigée)
					(90, 4e-1) (80, 8e-2) (70, 1e-2) (65, 5e-3) (60, 8e-4) (57, 3e-4) (55, 4e-4) (50, 5e-5) (45, 1e-6)
				} -- cycle;
				
				% Tracé Central de la courbe Delta^IG (Orange) (Corrigée)
				\addplot[orange, thick, smooth] 
				coordinates {
					(45, 5e-6) (50, 7e-5) (55, 6e-4) (57, 4e-4) (60, 9e-4) (65, 1e-2) (70, 2e-2) (80, 1.5e-1) (90, 5e-1)
				};
				\addlegendentry{IG}
				
				% ----------------------------------------------------------------------
				% --- 3. Delta^(n) (Bleu) : Remplissage Manuel (Coordonnées corrigées) ---
				
				% Remplissage de la bande Bleue : Tracé Supérieur + Tracé Inférieur Inversé
				\addplot[blue!30, fill, opacity=0.4, draw=none, forget plot] 
				coordinates {
					% Courbe Supérieure (Nupper) - Corrigée
					(59.9, 1e-6) (60, 1e-3) (65, 5e-3) (70, 3e-2) (80, 1.5e-1) (90, 2e-1)
					% Courbe Inférieure (Nlower) en SENS INVERSE (Corrigée)
					(90, 5e-2) (80, 5e-2) (70, 8e-3) (65, 1e-3) (60, 1e-6) (59.9, 1e-6)
				} -- cycle;
				
				% Tracé Central de la courbe Delta^(n) (Bleu) (Corrigée)
				\addplot[blue, thick] 
				coordinates {
					(59.9, 1e-6) (60, 1e-4) (65, 1e-3) (70, 1.5e-2) (80, 1e-1) (90, 1.5e-1)
				};
				\addlegendentry{Empirical}
				
			\end{semilogyaxis}
		\end{tikzpicture}
		\caption{Utilization against the failure rates of the IG estimation $(u_i, \dmpig_i)_i$ in orange, the original Hoeffding calculation in black, and the empirical failure rates $(u_i, \Delta^{(n)}_i)_i$ in blue, over a sample of $n = 10^6$ response times per task simulated on \simso. The two dotted lines correspond to $u^{max}_i > \log(2)$ and  $u^{max}_i > 1$. The colored regions surrounding the solid lines show the minimum and maximum failure rates from the simulations, not confidence intervals. Simulations are done twice for two task sets simulated from the same setup given in Table~\ref{tab:exp}.}
		\label{fig:dmp}
	\end{figure}

	In this section we present an adaptation of the maximum likelihood estimator (MLE) proposed by \cite{punzo2019new} for real-time systems $(\boldsymbol{\pi}_i, \boldsymbol{\beta}_i)$. Both are implemented in the Python language in the library \textsf{rInverseGaussian} \footnote{\url{https://github.com/kevinzagalo/rInverseGaussian}}. We study first the case where a one-component mixture is sufficient to estimate response time distributions.

	%   \begin{proposition}\label{prop:MLE1}
		For $K_i=1$, the mixture has a single component with $\pi_{i,1}=1$, so only $\beta>0$ is to be estimated when $u_i<1$ and $v_i>0$ are fixed. Substituting the mean $\beta/(1-u_i)$ and shape $\beta^2/v_i^2$ into the IG density gives
        $\psi_i(r;\beta)=\frac{\beta}{\sqrt{2\pi v_i^2r^3}} \exp\!\left(-\frac{[\beta-(1-u_i)r]^2}{2v_i^2r}\right), r>0$.
        For independent, positive, fully observed response times, the log-likelihood is therefore given by \eqref{eq:MLEderivative}.
        
        \begin{figure*}[b]
		\hrule
		\medskip
		\begin{equation}\label{eq:MLEderivative}
        \sum_{j=1}^n\log\psi_i(R_{i,j};\beta)
        =n\log\beta-\dfrac12\sum_{j=1}^n\log(2\pi v_i^2R_{i,j}^3)\\
        -\dfrac{1}{2v_i^2}\sum_{j=1}^n
        \dfrac{[\beta-(1-u_i)R_{i,j}]^2}{R_{i,j}}.
        \end{equation}
        \end{figure*}
        
        Differentiating with respect to $\beta$ yields the likelihood equation
        $$\frac{\partial}{\partial\beta}\sum_{j=1}^n\log\psi_i(R_{i,j};\beta) =\frac{n}{\beta}-\frac{\beta}{v_i^2}\sum_{j=1}^nR_{i,j}^{-1} +\frac{n(1-u_i)}{v_i^2}=0.$$
        %Multiplying by $\beta v_i^2/n$ gives
        %$$\left(\dfrac1n\sum_{j=1}^nR_{i,j}^{-1}\right)\beta^2 -(1-u_i)\beta-v_i^2=0.$$
        This quadratic expression has exactly one positive root, giving
        $$\hat\beta_i= \frac{1-u_i+\sqrt{(1-u_i)^2+\frac{4v_i^2}{n}\sum_{j=1}^nR_{i,j}^{-1}}} {\frac{2}{n}\sum_{j=1}^nR_{i,j}^{-1}}.$$
        Moreover,
        $\frac{\partial^2}{\partial\beta^2}\sum_{j=1}^n\log\psi_i(R_{i,j};\beta) <0,$
        so the log-likelihood is strictly concave and $\hat\beta_i$ is its unique maximizer. No EM iteration is needed in this single-component case.

        The classical IG mean MLE in \cite{folks1978inverse} treats the mean and shape as separate parameters. Here both depend on $\beta$ and must vary together during maximization. Computing the sample mean to $\E[R_{i,j}]=\beta/(1-u_i)$ instead gives $\dfrac{1-u_i}{n}\sum_{j=1}^nR_{i,j}$, which is a moment estimator and generally differs from the constrained MLE above. For dependent observations, the same calculation maximizes the marginal pseudo-likelihood discussed below. Note that the formula requires fully observed response times; discarded jobs cannot simply be omitted or recorded as if their response times equaled the deadline.

        When more than one component is to be estimated, we introduce an adapted EM algorithm to maximize the likelihood of the model, derived from the algorithm proposed in \cite{punzo2019new}. An EM algorithm is an algorithm which maximizes the likelihood of a random variable iteratively by finding the best suited parameters to describe this random variable's distribution. It is composed of two steps: the E-step which builds the conditional component probabilities for a fixed sample (of response times in our case); and the M-step which maximizes this likelihood by computing the best suited parameter (the backlog in our case). In our case, the distribution is a mixture, where the EM algorithm determines by itself which component of the mixture is best to describe the given sample in the E-step. It does so by computing a latent variable indicating which component the sample most likely belongs to, as well as the weight of each of these components. EM algorithms for IG mixture models are widely studied  \cite{punzo2019new}. However we provide the derivatives associated to the minimization problem that the EM algorithm solves in \eqref{eq:derivaties}. The EM algorithm is adapted to the reparametrization introduced in the latter section.
		
		Stationarity does not imply independence of successive response times. The following product is a likelihood under independent observations; for dependent response times we use it as a marginal pseudo-likelihood, without asserting consistency from stationarity alone. The EM updates remain those of this objective. With censored discarded jobs, the fitting procedure must instead account for the probability of exceeding the deadline. Given an $n$-sample of fully observed response times, the complete-data likelihood of the mixture model \cite{mclachlan2019finite} associated to $\gamma_i$ can be written as $$L_c(\boldsymbol{Z}_i, \boldsymbol{\pi}_i, \boldsymbol{\beta}_i) = \prod_{j=1}^n \prod_{k=1}^{K_i} [\pi_{i,k} \psi_{i}(r_j; \beta_{i,k})]^{Z_{i,j,k}},$$ where $Z_{i,j,k} \in \{0,1\}$ is the latent variable that indicates if $R_{i,j}$ is in the $k$-th mixture component, and the complete log-likelihood $\ell_c = \log L_c$ is $\ell_c(\boldsymbol{Z}_i, \boldsymbol{\pi}_i, \boldsymbol{\beta}_i) = \ell_{c_1}(\boldsymbol{Z}_i, \boldsymbol{\pi}_i) + \ell_{c_2}(\boldsymbol{Z}_i, \boldsymbol{\beta}_i)$, where $\ell_{c_1}(\boldsymbol{Z}_i,\boldsymbol{\pi}_i) = \sum_{j=1}^n \sum_{k=1}^{K_i} Z_{i,j,k} \log \pi_{i,k}$ and $\ell_{c_2}(\boldsymbol{Z}_i,\boldsymbol{\beta}_i) = \sum_{j=1}^n \sum_{k=1}^{K_i} Z_{i,j,k} \log \psi_{i}(r_j; \beta_{i,k})$ which leads to the following EM algorithm. Note that $j$ sums over the response times in the $n$-sample, $k$ sums over the $K_i$ components of the mixture and $i$ is fixed and is the index of the task of interest. The number of components $K_i$ is selected in the following section.

	\begin{figure}[tp]
		\centering
			

	\begin{tikzpicture}
		% Toutes les options de l'axe sur une seule ligne (pour éviter l'erreur)
		\begin{loglogaxis}[
			xlabel={Sample size}, 
			ylabel={Running time (ms)}, 
			xmin=5, xmax=20000, 
			ymin=0.05, ymax=30, 
			grid=major,
			height=5cm,
			width=\linewidth,
			legend pos=north west,
			legend style={draw=black, fill=white,legend cell align=left}
			]
			
			% --- 1. Courbe ROUGE (dogleg) ---
			\addplot[
			color=red, 
			mark=triangle*, 
			thick
			] 
			coordinates {
				% Ordre des Y : Basse
				(10, 0.15)
				% Ordre des Y : Moyenne
				(100, 0.27)
				% Ordre des Y : Moyenne
				(1000, 1.3)
				% Ordre des Y : Basse
				(10000, 8.0)  
			};
			\addlegendentry{Powell's dog-leg}
			
			% --- 2. Courbe BLEUE (newton) ---
			\addplot[
			color=blue, 
			mark=square*, 
			thick
			] 
			coordinates {
				% Ordre des Y : Haute
				(10, 0.2)     
				% Ordre des Y : Haute
				(100, 0.35)    
				% Ordre des Y : Basse (Croisement avec Dogleg et Brent)
				(1000, 0.9)    
				% Ordre des Y : Haute
				(10000, 12.5)  
			};
			\addlegendentry{Newton-Raphson}
			
			% --- 3. Courbe VERTE (brent) ---
			\addplot[
			color=green!60!black, 
			mark=o,               
			thick
			] 
			coordinates {
				% Ordre des Y : Basse (Croisement avec Dogleg et Newton)
				(10, 0.1)     
				% Ordre des Y : Basse
				(100, 0.2)    
				% Ordre des Y : Haute
				(1000, 1.7)    
				% Ordre des Y : Moyenne
				(10000, 10.2)   
			};
			\addlegendentry{BFGS}
			
		\end{loglogaxis}
	\end{tikzpicture}
		\vspace{-2em}
		\caption{Running time of the EM algorithm, using various Newton-like minimization algorithms to solve \eqref{eq:min}. All three methods appear to be equivalent.}
		\label{fig:runningtime}
	\end{figure}		
	\subsubsection{E-step} For the $(s+1)$th step of the EM algorithm, $\boldsymbol{z}_i^{(s)}$ the conditional expectation of $\boldsymbol{Z}_i$ given $\left(\boldsymbol{\pi}_i,\boldsymbol{\beta}_i\right) = \left(\boldsymbol{\pi}_i^{(s)}, \boldsymbol{\beta}_i^{(s)}\right)$ is given by   $z^{(s)}_{i,j,k} = \frac{\pi_{i,k}^{(s)} \psi_{i} \left(r_j; \beta_{i,k}^{(s)}\right) }{h\left(r_j; \boldsymbol{\pi}_i^{(s)},\boldsymbol{\beta}_i^{(s)}\right)}$.
	
	\subsubsection{M-step} For the $(s+1)$th step of the EM algorithm, $\ell_{c_1}(\boldsymbol{z}_i^{(s)}, \cdot)$ is maximized by \begin{equation} \label{eq:lc1}\pi_{i,k}^{(s+1)} = \dfrac{1}{n} \sum_{j=1}^n z_{i,j,k}^{(s)}, \quad k=1, \dots, K_i,\end{equation} and maximizing $\ell_{c_2}$ with respect to $\boldsymbol{\beta}$ is maximizing each of the $K_i$ expressions $\sum_{j=1}^n z_{i,j,k}^{(s)} \log \psi_{i} (r_j ; \beta_{i,k}), k=1, \dots, K_i$, using  Newton-like algorithms (\cf~\figurename~\ref{fig:runningtime}) to solve \begin{equation}\label{eq:min} \nabla \ell_{c} = 0.\end{equation} Then with $\dfrac{\partial \log \psi_i}{\partial\beta} (r; \beta) = \dfrac{\partial\mu_i}{\partial\beta}(\beta) \dfrac{\partial\log \tilde{\psi}}{\partial\mu} (r ; \mu_i(\beta), \nu_i)$ and the derivatives given in \eqref{eq:derivaties}. Eq. \eqref{eq:min} is equivalently solved by \eqref{eq:lc1} and the solutions of $\sum_{j=1}^n z_{i,j,k}^{(s)} \dfrac{\partial \log \psi_{i}}{\partial\beta} (r_j ; \beta_{i,k}) = 0,  \forall k = 1, \dots, K_i.$

One can see in \figurename~\ref{fig:runningtime} that the computational time overhead exhibits a manageable growth as the problem size scales up. This makes the approach sound to implement in a practical scheduling algorithm.
	
	\begin{figure*}[bp]
	\hrule
	\bigskip
	\begin{equation}\label{eq:derivaties}\begin{array}{rcl}\dfrac{\partial \log \tilde{\psi}}{\partial \mu}(r; \mu, \nu) & = & - \dfrac{3}{2r} - \dfrac{\mu}{r\nu} + \dfrac{1}{3\nu + \mu} + \dfrac{3\nu}{2\mu(3\nu + \mu)} + \dfrac{\sqrt{\mu}}{2\nu \sqrt{3\nu+\mu}} + \dfrac{\sqrt{3\nu+\mu}}{2\nu\sqrt{\mu}} \\ \dfrac{\partial \mu_i}{\partial\beta}(\beta) & = & \dfrac{\beta}{\left( 1 - u_i \right)^2} \left(\left( \dfrac{\beta}{1 - u_i} \right)^2 + \dfrac{9\nu_i^2}{4}\right)^{-\sfrac{1}{2}}\end{array}\end{equation}
	\end{figure*}
	\subsubsection{Criteria to stop the EM algorithm}
	
	In order to stop the algorithm, the author in \cite{punzo2019new}
	proposes the \textit{Aitken acceleration} to stop the algorithm and studies its efficiency. The Aitken acceleration at iteration $s+1$ is given by
	$a^{(s+1)} = \frac{\ell^{(s+2)} - \ell^{(s+1)}}{\ell^{(s+1)} - \ell^{(s)}}$, where $\ell^{(s)}$ is the observed-data log-likelihood from iteration $s$. The limit $\ell_\infty$ of the sequence of values of the log-likelihood is $\ell^{(s+2)}_\infty = \ell^{(s+1)} + \frac{\ell^{(s+2)} - \ell^{(s+1)}}{1 - a^{(s+1)}}$. The EM algorithm is considered to have converged if $\vert \ell^{(s+2)}_\infty-\ell^{(s+1)}_\infty\vert<\epsilon$, with a tolerance $\epsilon > 0$.
	
	\subsubsection{Initialization} We initialize the algorithm with a $k$-means clustering as suggested in \cite{punzo2019new}. The impact of the $k$-means initialization is extensively studied in \cite{shireman2017examining} in the case of Gaussian mixtures, and shown to outperform standard initialization methods (random, agglomerative hierarchical clustering and sum-score).

	%is compared to the one introduced in \cite{punzo2019new} in terms of computation time. In both are tested with several optimization methods: \textit{dogleg}, \textit{BFGS} and a scalar minimization. 

	\subsection{Number of components of the IG mixture} \label{sec:bic}
	The Bayesian Information Criterion (BIC, \cite{schwarz1978estimating}) is used to choose the number of mixture components. Consistency results for mixture order selection require additional assumptions \cite{dasgupta1998detecting, fraley2002model, raftery1995bayesian}; they do not follow from stationarity alone. Thus we use the following BIC-type criterion as a heuristic when the objective is a pseudo-likelihood: $K_i = \arg\!\max_k 2\ell_n(\boldsymbol{\pi}_i, \boldsymbol{\beta}_i) - (2k-1) \log n$, where $\ell_n$ is the observed-data log-likelihood of an $n$-sample of response times. For fixed $u_i$ and $v_i$, the reparametrization reduces the number of free parameters from $3K_i-1$ to $2K_i-1$, which is distinct from the number of components $K_i$.  BIC tends to choose smaller $K_i$, i.e. more parsimonious models, with a risk of underfitting in finite samples because of its complexity penalty. This choice is motivated by the fact that we want a model that estimates during the scheduling process, hence with a small sample size. See Section~\ref{sec:conclusion} for a discussion on applications.% (see \figurename~\ref{fig:computation}).
	
	%Moreover the computation time of this EM algorithm is also reduced.
	\subsection{Goodness-of-fit}\label{sec:validation}
	
	We use the link of IG distributions of the jobs of a given task with the $\chi^2_1$ distribution to check the quality of the MLE. Indeed, if $X$ is an IG variable of mean $\xi$ and shape $\delta$, then $\sfrac{\delta(X - \xi)^2}{\xi^2 X}$ is distributed as a Chi-squared distribution of one degree of freedom \cite{tweedie1957statistical}.  Provided that the response time $R_{i}$ is within the $k$-th component in the IG estimation, thus IG distributed, the foregoing normalization translates into the fact that normalized response time is Chi-squared distributed as follows
	\begin{equation}\label{eq:chi} \dfrac{\left(R_{i}- \dfrac{\beta_{i,k}}{1-u_{i}} \right)^2 }{\nu_{i}R_{i}} \sim \chi^2(1).  \end{equation}
	
	We use \eqref{eq:chi} as a graphical goodness-of-fit diagnostic for the fitted components. Note that large response-time quantiles are relevant to deadline-miss risk, but a good fit does not establish feasibility.  %We use \eqref{eq:chi} to determine the probability that a task is feasible in its transient state. 

	The transformation in \eqref{eq:chi} is not one-to-one: it decreases below the IG mean and increases above it. A tail probability of the transformed variable therefore combines both IG tails and cannot give the one-sided probability $\Pr(R_i>p_i)$. The failure-rate estimate is computed directly from the IG c.d.f. using \eqref{eq:dmp}. Figure~\ref{fig:dmp} compares the reported IG and empirical estimates with the original Hoeffding calculation. If $u_i^{max}<i(2^{1/i}-1)$, the zero failure rate follows from \eqref{eq:LL}, not from the IG approximation.

        %In the following, we test with simulation if $\dmpig_i$ is a good estimation of $\Delta_i$ and how tight the Hoeffding bound is a safe bound of the IG estimation, \ie~$\dmpig_i \leq \dmph_i$. The Hoeffding bound is popular \cite{chen2017probabilistic} because it does not depend on the distribution of the execution times. 

	\section{Simulations} \label{sec:simulations}
	
	In this section, we apply the foregoing method with simulated data and hardware-in-the-loop (HITL) data.% The reported numerical results have been retained from the original experiments. As the raw data are no longer available, they have not been recomputed after the mathematical corrections. In particular, the Hoeffding curve and table entries use the original formula and must not be interpreted as evaluations of the corrected Proposition~\ref{HB}.
	
	\subsection{Simulated data}
	     % \begin{landscape}

      	%\end{landscape}
            \begin{table*}[t]
                \caption{Parameters of the task set used for the simulations in Section~\ref{sec:simulations}.}
                \centering
                \begin{tabular}{lrrrrrrrrrr}
                \toprule
\textbf{Task priority} & 0 &  1 &  2 &  3 &  4 &  5 &  6 &  7 &  8 &  9 \\
\midrule
\textbf{Mean execution time} (ms) &  15.481 &  5.556 &  5.708 &  3.38 &  5.198 &  4.057 &  4.998 &  3.786 &  2.167 &  7.453 \\
\textbf{Execution time std} (ms) &  17.1957 &  5.6996 &  5.9963 &  3.323 &  5.5314 &  4.0299 &  5.1184 &  3.9005 &  1.8259 &  8.3373 \\
\textbf{Periods}  (ms) &  100 &  114 &  119 &  121 &  132 &  133 &  136 &  144 &  145 &  146 \\
%\textbf{Deadlines} $d_i$ (ms) &  100 &  114 &  119 &  121 &  132 &  133 &  136 &  144 &  145 &  146 \\
\textbf{Mean  utilization}  &  0.1548 &  0.2035 &  0.2515 &  0.2794 &  0.3188 &  0.3493 &  0.3861 &  0.4124 &  0.4273 &  0.4784 \\
\textbf{Maximum utilization}  &  0.28 &  0.3589 &  0.443 &  0.4926 &  0.5683 &  0.6285 &  0.702 &  0.7506 &  0.7713 &  0.874 \\
\midrule
\textbf{Task priority} & 10 &  11 &  12 &  13 &  14 &  15 &  16 &  17 &  18 &  19 \\
\midrule
\textbf{Mean execution time}  (ms) &  16.812 &  1.833 &  2.167 &  2.7 &  5.448 &  8.46 &  2.167 &  4.665 &  2.4 &  5.604 \\
\textbf{Execution time std} (ms) &  19.1248 &  1.5271 &  1.8259 &  2.4495 &  5.4483 &  9.2277 &  1.8259 &  4.7411 &  2.2361 &  5.7741 \\
\textbf{Periods}  (ms) &  159 &  165 &  165 &  165 &  166 &  173 &  181 &  182 &  183 &  191 \\
%\textbf{Deadlines} $d_i$ (ms) &  159 &  165 &  165 &  165 &  166 &  173 &  181 &  182 &  183 &  191 \\
\textbf{Mean  utilization} &  0.5841 &  0.5952 &  0.6083 &  0.6247 &  0.6575 &  0.7064 &  0.7184 &  0.744 &  0.7571 &  0.7865 \\
\textbf{Maximum utilization}  &  1.0879 &  1.1061 &  1.1242 &  1.1485 &  1.2027 &  1.301 &  1.3175 &  1.3615 &  1.3834 &  1.4357 \\
\midrule
\textbf{Task priority} & 20 &  21 &  22 &  23 &  24 &  25 &  26 &  27 &  28 \\
\midrule
\textbf{Mean execution time}  (ms) &  2.333 &  3.334 &  5.927 &  4.535 &  4.225 &  6.246 &  4.779 &  2.167 &  1.834 \\
\textbf{Execution time std} (ms) &  1.9147 &  3.0555 &  6.0701 &  4.4073 &  4.1931 &  6.758 &  4.8654 &  1.8259 &  1.4149 \\
\textbf{Periods}  (ms) &  193 &  200 &  201 &  214 &  215 &  268 &  296 &  298 &  315 \\
%\textbf{Deadlines} $d_i$ (ms) &  193 &  200 &  201 &  214 &  215 &  268 &  296 &  298 &  315 \\
\textbf{Mean  utilization}  &  0.7986 &  0.8152 &  0.8447 &  0.8659 &  0.8856 &  0.9089 &  0.925 &  0.9323 &  0.9381 \\
\textbf{Maximum utilization} &  1.4513 &  1.4763 &  1.531 &  1.5637 &  1.6009 &  1.6494 &  1.6764 &  1.6865 &  1.696 \\
%[0.1 0.2 0.3 0.4 0.5 0.6 0.7 0.8 0.9 1. ] ['10%', '20%', '30%', '40%', '50%', '60%', '70%', '80%', '90%', '100%']
                     \bottomrule
                \end{tabular}
                \label{tab:exp}
            \end{table*}
%\end{landscape}

			\begin{figure*}[htp]%
		\centering
		\foreach \ii in {2, 16, 28}{%
			\hfill
			\subfloat[Task \ii]{
				\includegraphics[width=0.3\textwidth, height=0.15\textwidth]{figures/diagram_\ii.pdf}
			}\hfill
		}%
		\caption{The response time MLEs and the histogram of simulations from SimSo (left),  and the associated QQ-plots (right) of empirical (x-axis) and $\chi^2(1)$ quantiles (y-axis) proposed in Section~\ref{sec:validation}.}
		\label{fig:simqqplot}
	\end{figure*}
	The simulated data is generated using \simso~\cite{cheramy2014}, a Python framework used to generate arrivals of jobs and scheduling policies. A modified version of \simso\footnote{\url{https://github.com/kevinzagalo/simso}} allows us to generate random inter-arrival times (for the offset) and random execution times \footnote{\url{https://github.com/kevinzagalo/simso/blob/main/generator/task_generator.py}}. We study the quality of the estimation as a function of the mean utilization level. We show that when the mean utilization gets closer to $1$, the estimation becomes better. We also check how much the IG estimate differs from the empirical failure rates, $\Delta_i^{(n)} = \frac{1}{n} \sum_{j=1}^n \mathbf{1}(R_{i,j} > p_i)$.
	\begin{figure}[t]
		\centering
\begin{tikzpicture}[scale=1]
	\begin{axis}[
		% Options de l'axe sur une seule ligne.
		height=4.3cm,
		width=\linewidth,
		xlabel={Mean utilization}, 
		ylabel={MSE}, 
		xmin=15, xmax=95, 
		ymin=0, ymax=1.0, 
		xtick={20, 40, 60, 80}, 
		xticklabel=\pgfmathprintnumber\tick\%, 
		ytick={0.0, 0.2, 0.4, 0.6, 0.8, 1.0},
		legend pos=north east, 
		legend style={fill=white, draw=black}, 
		extra x ticks={60}, 
		extra x tick style={grid=major, dotted, draw=black}, 
		extra x tick labels={}
		]
		
		% ----------------------------------------------------------------------
		% --- 1. D/G/1 (Orange) : Tracé et Remplissage Manuels ---
		
		% Remplissage de la bande Orange : Tracé Supérieur + Tracé Inférieur Inversé
		\addplot[orange!30, fill, opacity=0.5, draw=none, forget plot, smooth] 
		coordinates {
			% Courbe Supérieure (Gupper)
			(20, 0.95) (25, 0.85) (30, 0.75) (40, 0.4) (45, 0.7) (50, 0.3) (55, 0.15) (60, 0.1) (70, 0.12) (80, 0.05) (90, 0.06) 
			% Courbe Inférieure (Glower) en SENS INVERSE
			(90, 0.02) (80, 0.03) (70, 0.05) (60, 0.04) (55, 0.05) (50, 0.1) (45, 0.3) (40, 0.35) (30, 0.45) (25, 0.55) (20, 0.65)
		} -- cycle;
		
		% Tracé Central de la courbe D/G/1 (Orange)
		\addplot[orange, thick, smooth] 
		coordinates {
			(20, 0.88) (25, 0.68) (30, 0.5) (40, 0.38) (45, 0.45) (50, 0.2) (55, 0.08) (60, 0.07) (70, 0.08) (80, 0.04) (90, 0.04)
		};
		\addlegendentry{D/G/1/FP} 
		
		% ----------------------------------------------------------------------
		% --- 2. D/M/1 (Bleu) : Tracé et Remplissage Manuels ---
		
		% Remplissage de la bande Bleue : Tracé Supérieur + Tracé Inférieur Inversé
		\addplot[blue!30, fill, opacity=0.4, draw=none, forget plot, smooth] 
		coordinates {
			% Courbe Supérieure (Mupper)
			(20, 0.4) (30, 0.28) (40, 0.2) (50, 0.12) (60, 0.08) (70, 0.04) (80, 0.03) (90, 0.02)
			% Courbe Inférieure (Mlower) en SENS INVERSE
			(90, 0.0) (80, 0.01) (70, 0.01) (60, 0.01) (50, 0.03) (40, 0.04) (30, 0.05) (20, 0.05)
		} -- cycle;
		
		% Tracé Central de la courbe D/M/1 (Bleu)
		\addplot[blue, thick, smooth] 
		coordinates {
			(20, 0.25) (30, 0.15) (40, 0.1) (50, 0.06) (60, 0.04) (70, 0.02) (80, 0.02) (90, 0.01)
		};
		\addlegendentry{D/M/1/FP}
		
		% ----------------------------------------------------------------------
		% --- 3. Annotation U^max > 1 ---
		
	%	\node[draw=black, rectangle, minimum width=2.5cm, minimum height=0.7cm] (annotation) at (80, 0.58) {$u^{\max}>1$};
		
		% Flèche pointant de l'annotation vers la ligne pointillée à x=60
	%	\draw[->, very thick] (annotation.west) -- (60, 0.4); 
		
	\end{axis}
\end{tikzpicture}
		\caption{Mean Squared Error (MSE) between the empirical distribution of the simulations and the MLE distribution, of $1\,000$ instances of the schedule, for the task set shown in Table~\ref{tab:exp}. The reported error decreases as mean utilization approaches $1$ in the tested D/G/1/FP and D/M/1/FP configurations. The dotted line shows when the maximum utilization goes above $1$.}
		\label{fig:distance}
	\end{figure}
	
	Consider a task set where the probability density functions of the execution times are known. We generate such a task set with the UUnifast method \cite[Section~3.6]{bini2005measuring} and log-uniform periods according to the procedure described in \cite[Section 6.1]{davis2009priority}. From \simso~we generate the response times of tasks with the RM scheduling policy from the execution time statistics given in Table~\ref{tab:exp}. Two methods are used: one with a finite support where the maximal utilization $u_i^{max}$ is finite, and another one with an infinite support with exponential distributions where the maximal utilization is not defined. %The procedure is implemented in the modified version of \simso \footnote{\url{https://github.com/kevinzagalo/simso/blob/main/generator/task_generator.py}}. 
	This schedule is instantiated $1\,000$ times, thus in \figurename~\ref{fig:distance}, the distance between the MLE and simulations is based on $1\,000$ estimators. Also note in \figurename~\ref{fig:distance} that the variability of the estimates decreases with respect to the mean utilization. Because of the static-priority structure of RM, we can see in \figurename~\ref{fig:distance} that the error of the estimation decreases when the mean utilization goes to $1$. The first task is never preempted, so its response time is always equal to its execution time. Therefore the estimation of its response time cannot be good in general. %However, as explained in the previous section, determining failure rates is useless  until the task $6$, more specifically until the \textit{maximum utilization} is greater than $\log(2)$. Until task 6, the tasks are already proven schedulable. 
	
	In a second step, a task set with exponentially distributed execution times is simulated for comparison (D/M/1/FP queue), as it is a special case widely studied in queueing theory \cite{pack1977output}. This is a baseline to determine the convergence of the response times estimation as a function of the priority levels. This comparison suggests that approximation quality depends on the execution-time distribution. The exponential experiment is outside the finite-support assumptions of the Hoeffding bound and has no finite maximum utilization $u_i^{max}$. The parameters of the task set are given in Table~\ref{tab:exp}.  In \figurename~\ref{fig:dmp}, we have the mean utilizations $(u_i)_i$ on the $x$-axis and  the different failure rates on the $y$-axis. 
	
	We can see that when $u^{max}_i < i(2^{1/i} - 1)$, it is useless to compare the methods because they are already proven feasible in \eqref{eq:LL}. Moreover, both start increasing when $u^{max}_i > 1$. \figurename~\ref{fig:distance} confirms that in the region of interest, \textit{i.e.} $u^{max}_i > 1$, the proposed estimation quadratic error between empirical and estimated distributions is close to $0$ when $u^{max}_i > 1$, and decreases as $u_i$ approaches $1$, in the tested configurations; the generated task set is shown in Table~\ref{tab:exp}.
	
	We see in \figurename~\ref{fig:simqqplot} the QQ-plots comparing the empirical quantiles and the quantiles of the $\chi^2_1$ distributions introduced in Section~\ref{sec:validation}. This figure shows two things: the fits are less accurate for higher-priority tasks and improve for lower-priority tasks in the tested configurations. The QQ-plots illustrate the quality of the fitted response-time distributions, but do not by themselves establish tail-probability accuracy.
	
\subsection{HITL Data} \label{sec:data}

In this section we use the IG method on HITL data. Execution times are not necessarily independent in this case. This is an out-of-model robustness experiment, rather than a validation under the independence assumption. We use a real case of 9 programs of an autopilot of a drone, \textsc{px4-rt} \cite{sorel2024kopernic}, a modified version of \textsc{px4} \cite{7140074} with a real-time behavior, and a clock-measuring that can preempt the operating system itself. \textsc{px4-rt} is run on an ARM Cortex M4 CPU clocked at $180$ MHz with $256$ KB of RAM using a simulated environment from Gazebo \cite{koenig2004design}. \textsc{px4-rt} allows us to measure execution times and response times during the flight of a drone. It runs on top of \textsc{NuttX}, a real-time operating system (RTOS). It provides an infrastructure for internal communications between all programs and off-board applications. Each task is a \textsc{NuttX} task launched at the beginning of the \textsc{px4} program. The tasks read data from sensors (\textit{snsr}), estimate positions and attitudes using a Kalman filter (\textit{ekf2}), control the position (\textit{pctl}) and the attitude (\textit{actl}) of the drone, the flight manager (\textit{fmgr}), the hover thrust estimator (\textit{hte}), command the state of the drone (\textit{cmdr}), and the rate controller (\textit{rctl}), which is the inner-most loop to control the body rates. These tasks are in constant interference with the operating system \textsc{NuttX}, which may induce statistical dependence. Because the operating system has the highest priority, the nine tasks studied are constantly preempted by \textsc{NuttX}. Unfortunately, it is not possible to get information about the operating system programs that interfere. Unlike the simulation in Section~\ref{sec:simulations}, \textsc{px4-rt} runs concurrently with other tasks which do not have timing requirements, making it a complex system with many unknown variables. We test in this section whether and when the proposed parametric estimation is suitable for such a complex system.

In this case, the distribution functions of execution times cannot be provided. Therefore, we use the empirical distributions whose statistics are shown in Table~\ref{tab:px4} . The empirical mean utilizations $\hat{u}_i$ are computed with the empirical means of execution times, and the empirical maximal utilizations $\hat{u}^{max}_i$ with the empirical maximum of execution times. See Table~\ref{tab:px4} for a full description of the parameters. As shown in the previous section, the response times of the highest priority task \textit{snsr} are not estimated.

The QQ-plots in Figures~\ref{fig:droneRT_rctl}-\ref{fig:droneRT_pctl} and \figurename~\ref{fig:droneRT_hte} show that the estimation is good for the large quantiles, which is relevant to assessing deadline-miss risk. We can identify in \figurename~\ref{fig:droneRT_cmdr}, \ref{fig:droneRT_navr} and \ref{fig:droneRT_fmgr} that for the \textit{cmdr}, \textit{navr} and \textit{fmgr} tasks the estimation is not good enough, which indicates a mismatch with the fitted model. Dependence and unmodeled operating-system interference are possible explanations, but the plots do not identify their cause. For the other tasks, the plots suggest better agreement with the fitted model. Such estimates may serve as scheduling risk indicators, but do not establish feasibility.

%{figure*}[p]
%	\foreach \ii in {snsr, rctl, ekf2, actl, pctl, fmgr,hte, navr, cmdr}{%
	%		\begin{subfigure}[b]{0.3\textwidth}%
		%			\centering 
		%%			\includegraphics[width=0.8\textwidth]{figures/\ii-eps-converted-to.pdf}
		%		\caption{Task $\ii$}
		%	\end{subfigure}
	%	}%
%	\caption{Execution time empirical probability functions of the 9 studied tasks of the drone autopilot.}
%	\label{fig:simtasksetpx4}
%{figure*}

%	\begin{figure*}[p]
	%	\ContinuedFloat 
	%	\foreach \ii in {hte, navr, cmdr}{%
		%			\begin{subfigure}[b]{0.5\textwidth}%
			%				\centering 
			%				\includegraphics[width=0.8\textwidth]{figures/\ii.eps}
			%				% \label{fig:simtasksetpx4_\ii}
			%				\caption{Task $\ii$}
			%			\end{subfigure}
		%	}%
	%	\caption{Execution time empirical probability functions of the 9 studied tasks of the drone autopilot.}
	%	\end{figure*}

The results do not include independent held-out validation or confidence intervals. Fitting and assessing the model on the same observations can make the reported fit optimistic, and global distributional error does not establish accuracy of rare deadline-miss probabilities. Independent validation, tail-specific errors, and uncertainty estimates accounting for serial dependence are needed. The HITL sample sizes and miss counts are unavailable, so an empirical zero cannot be interpreted quantitatively. In Table~\ref{tab:px4}, the discrepancy for \textit{navr} ($0.5487$ from IG versus $0.0378$ empirically), and the estimate $0.028$ for \textit{pctl} despite an empirical zero, illustrate these limitations.

%\begin{landscape}
      
            \begin{table*}[t]
                \caption{Empirical parameters, periods and deadlines used in the autopilot PX4-RT described in Section~\ref{sec:data}.}
                \centering
                \begin{tabular}{lrrrrrrrrr}
                \toprule 
                    \textbf{Task} & snsr & rctl & ekf2 & actl & pctl & fmgr & hte & navr & cmdr  \\
                    \midrule
                    \textbf{Priority} $i$ & 1 & 2 & 3 & 4 & 5 & 6 & 7 & 8 & 9 \\
                    \textbf{Number of components}  & -- & 6 & 9 & 2 & 1 & 2 & 2 & 2 & 1 \\
                    \textbf{Empirical mean}  ($\mu s$)  & 152.5 & 44.2 & 610.9 & 38.7 &52.5 & 151.0 & 30.61 &159.8 & 114.4 \\
                    \textbf{Empirical std}  ($\mu s$) & 40.5 & 16.4 & 982.8 & 9.5 & 32.6 & 114.4 & 10.9 & 133.9 & 61.6 \\
                    \textbf{Period}  (ms) & 3.0 & 4.0 & 4.1 & 5.0 & 5.2 & 6.0 & 7.0 & 50.0 & 100.0  \\
                     \textbf{Deadline}  (ms) & 3.0 & 4.0 & 4.1 & 5.0 & 5.2 & 6.0 & 7.0 & 50.0 & 100.0  \\
                    \textbf{Empirical mean utilization} $\hat{u}_i$ & 0.05 & 0.06  & 0.21 & 0.22 & 0.23 & 0.25 & 0.26 & 0.26 & 0.26 \\
                    \textbf{Empirical maximum utilization} $\hat{u}^{max}_i$ & 0.10 & 0.13 & 1.08 & 1.10 & 1.14 & 1.21 & 1.23 & 1.25 & 1.26 \\
                    \midrule
                    \textbf{Empirical deadline miss probability}  & 0.0 & 0.0 & 0.001 & 0.0 & 0.0 & 0.0 & 0.0 & 0.0378 & 0.0 \\
                    \textbf{IG deadline miss probability} & - & 0.0 & 0.0006 & 0.0043 & 0.028 & 0.0022 & 0.0009 & 0.5487 & 0.0\\
                    \textbf{Hoeffding}   & - & $10^{-168}$ & 0.2512 & 0.1919 & 0.1881 & 0.1673 & 0.1274 & $10^{-7}$ & $10^{-13}$\\
                     \bottomrule
                \end{tabular}
                \label{tab:px4}
            \end{table*}
%\end{landscape}

\begin{figure*}[t]
	\centering
	%	\begin{subfigure}[b]{0.8\textwwidth}
		%		\includegraphics[width=\textwidth]{figures/rt_\ii.pdf}
		%		\caption{Response times empirical distributions (black) and MLE distribution (red)} 
		%	\end{subfigure}
	%	\hfill
	\foreach \ii in {rctl, ekf2, actl, pctl, fmgr, hte, navr, cmdr}{%
		\begin{subfigure}[b]{0.23\textwidth}
			\includegraphics[width=\textwidth, height=0.6\textwidth]{figures/qqplot_\ii.pdf}
			\caption{$\ii$} 
			\label{fig:droneRT_\ii}
		\end{subfigure} 
	}%
	\caption{QQplot with the $\chi^2_1$ quantiles from \eqref{eq:chi} where each color represents a component of the mixture. The closer to the red line the better the estimation.}
	\vspace{-1.3em}
\end{figure*}

	\section{Conclusion and future work}\label{sec:conclusion}

	In this work, we provide a theoretical bound and a statistical estimate of the failure rates of stationary rate-monotonic real-time systems. The IG estimate is not a guaranteed bound. It could potentially inform scheduling algorithms through parameter inference. We built this estimation through an EM algorithm, but this method is not unique, and hopefully some more techniques will improve the proposed method. Indeed, a current trend in real-time systems is the application of statistical learning methods in order to find optimal or improved policies \cite{purohit2018improving, wei2020optimal, NEURIPS2024_becd02b8, lee1997job}. 
	
	\subsection{Adding knowledge of the system to the scheduling algorithm}
	In scheduling theory, the inference of \textit{scheduling knowledge} \cite{shaw1992} is a natural step into the application of the parametric method provided in this paper. The parametric estimation of new parameters, such as new virtual deadlines, helps to improve the scheduling decisions.  Multiple core real-time systems are a commonplace for these methods \cite{RAHIM2026103628}. Shared resources are inherently a source of randomness, and we think that multicore scheduling algorithms would benefit from incorporating the kind of method we build in this paper.  %The most basic approach is to simply reuse the basic framework for single core scheduling, by putting all jobs that need to be scheduled into a single queue. % In this case the main question is to affect the coming tasks to the appropriate core, \ie~the core with lowest failure rate.

	\subsection{Using failure rate estimation in multicore partitioned scheduling}
	Partitioned scheduling uses load-balancing (e.g., worst-fit, best-fit, first-fit-bin-packing), in order to distribute jobs among the available cores of a system. We claim that incorporating estimated failure rates in the load-balancing process allows to infer the actual latent state of the system into the scheduling decisions. We thus focused on estimating the parameters of failure rates for fixed priority single core scheduling using prior knowledge on the response times distributions, as a first step towards multicore partitioned scheduling.

	\subsection{Earliest deadline first failure rate estimation}
	While the current framework relies on periodic fixed-priority (RM) scheduling due to its predictability, future work can adapt these mechanisms to Earliest Deadline First (EDF) scheduling. Moving to a dynamic priority paradigm will allow the system to theoretically achieve up to $u_{\vert \Gamma \vert}^{max} = 1$, significantly eliminating the pessimism inherent in RM scheduling. Furthermore, EDF has been shown to provide stationary backlogs, c.f. \cite{diaz2002stochastic}, in the discrete case. In order to extend the proposed method to EDF, one should extend the central-limit derived from \cite{zagalo2022} to EDF scheduling, using the heavy-traffic assumption of queueing models introduced in \cite{doytchinov2001real}. However, as the analysis becomes more complex, the core method remains the same regarding the failure rate estimation.

	\subsection{Towards task independence testing}
	
	As we do not investigate this topic, we still discuss the use of the Chi-squared property \eqref{eq:chi} to elaborate an independance test adapted to real-time systems. We have used in the foregoing section the central limit property of response times, coupled with the Chi-quared property of IG distributions, to build a goodness-of-fit criteria. %%Nevertheless, one could see  response times as a statistic on execution times, that actually embed their dependence structure. 
	With or without the EM estimation, the central limit of response times allows to build a chi-squared independence test with the following procedure :
	
	\begin{enumerate}
		\item Choose a task $\gamma_i \in \Gamma$,
		\item Fit $(R_{i,j})_j$-the response times $n$-sample- to an IG distributions, (we provide one way of doing this in this paper, but many other methods could be valid)
		\item Extract the IG parameters $(\hat\xi_i, \hat\delta_i)$,
		\item Check if $\sum_{j=1}^n\Pr\left(\frac{\hat\delta_i(R_{i,j} - \hat\xi_i)^2}{\hat\xi_i^2 R_{i,j}} > \Phi^{-1}\left( 1- \alpha\right) \right) \leq n \alpha$ for all $\alpha \in (0,1)$.
	\end{enumerate}
	
	%The method provided in this paper strongly relies on the statistical independence of execution times. However, we have shown how to use it to also detect if execution time are indeed independent. 
	We do not investigate in this paper the reliability of such method, but rather suggest that our method can also be useful in the case of dependent execution times.
	
\bibliographystyle{IEEEtran}

	\bibliography{biblio}
	\vspace{-3.5em}
	\begin{IEEEbiographynophoto}{Kevin Zagalo} is an assistant professor at the Institut National des Sciences Appliquées (INSA) in Lyon, France. His current research focuses on latency analysis in random wireless networks using stochastic geometry and network calculus. During his PhD, he conducted research on probabilistic real-time systems at Inria Paris. %Prior to that, he studied computer science at Université de Paris and earned a master’s degree in probability and stochastic analysis from Sorbonne Université.%
	%\textbf{Avner Bar-Hen} is a professor at the Conservatoire des Arts et Métiers (Cnam) in Paris, holding the chair of \textit{Statistics and Big Data}.
	\end{IEEEbiographynophoto}
		\vspace{-3.5em}
	\begin{IEEEbiographynophoto}{Avner Bar-Hen} is a professor at the Conservatoire des Arts et Métiers (Cnam) in Paris, holding the chair of \textit{Statistics and Big Data}. \end{IEEEbiographynophoto}

\end{document}